\documentclass[preprint,prl,aps]{revtex4}

\input epsf
\usepackage{graphicx}

\begin{document}
\draft
\title{NUCLEON ELECTROMAGNETIC FORM FACTORS}
\author{Haiyan Gao}
\address{Triangle Universities Nuclear Laboratory, and Department of Physics, 
Duke University, Durham, North Carolina 27708, U.S.A.}


\begin{abstract}

The nucleon electromagnetic form factors have been 
studied in the past extensively from unpolarized electron scattering 
experiments. With the development in polarized beam, recoil polarimetry, 
and polarized target 
technologies, polarization experiments have provided more precise 
data on these quantities.
In this talk, I review recent experimental 
progress on this subject.

\end{abstract}
\maketitle
\section{Introduction}	

The electromagnetic form factors of the nucleon  
are fundamental quantities describing
the distribution of charge and magnetization within nucleons.
Quantum Chromodynamics (QCD) is the theory of strong interaction in terms
of quark and gluon degrees of freedom. While QCD has been extremely well
tested in the high energy regime, where perturbative QCD is applicable,
understanding confinement and hadron structure in the non-perturbative
region of QCD remains challenging.
Knowledge of the internal structure of protons and neutrons in
terms of quark and gluon degrees of freedom is not only essential for testing 
QCD in the confinement regime, but it also provides a basis for 
understanding more complex, strongly interacting matter 
at the level of quarks and gluons.

\section{Proton Electromagnetic Form Factors}

The proton electric ($G^p_E$) and magnetic ($G^p_M$) form factors 
have been studied extensively in the
past from unpolarized electron-proton ($ep$) elastic scattering using the  
Rosenbluth separation technique \cite{rosenbluth}.
New data from polarization transfer experiments \cite{mjones,gayou}, which 
measure this ratio directly with unprecedented precision,
show very intriguing behavior at higher $Q^2$. 
The form factor ratio, 
${\frac{\mu G^p_E}{G^p_M}}$ drops to approximately 0.5 at a $Q^2$ value 
above 3 (GeV/c)$^2$, and to approximately 0.3 at the highest measured 
$Q^2$ value ($\sim$ 5.5 (GeV/c)$^2$).
No such dramatic behavior in this ratio had been observed 
from unpolarized cross section measurements. 

\begin{figure}[htbp]
\includegraphics[width=3.0in]{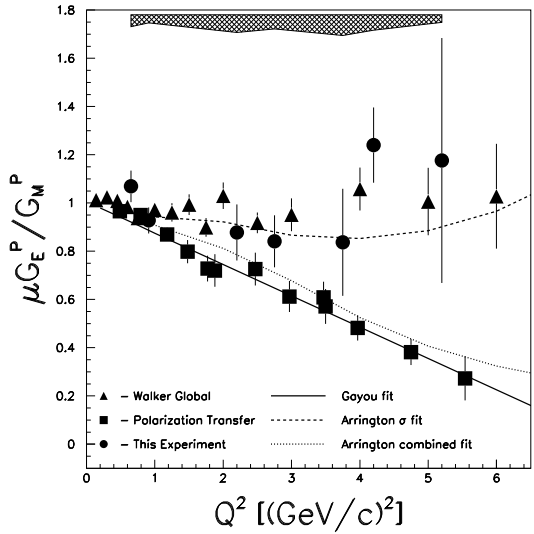}
\caption{\footnotesize\it Proton electric to magnetic form factor 
ratio as a function of $Q^2$. Data from JLab recoil proton polarization 
measurements~\cite{mjones,gayou} are shown as solid squares, 
the new Jefferson Lab data from Rosenbluth separation~\cite{hallc} 
are shown as solid circles
together with the error band representing the absolute uncertainty 
due to the scattering angle uncertainty. The SLAC data are shown as solid 
triangles~\cite{slac}. The dashed line is a reanalysis~\cite{ja} of 
global unpolarized data~\cite{slac}. The dotted line~\cite{ja2} 
is a fit by combining 
the cross-section data and the recoil polarization data.}
\label{fig:gepgmp}
\end{figure}

Fig.~\ref{fig:gepgmp} shows the proton electric to magnetic form factor ratio
as a function of $Q^2$ from recoil proton polarization measurements at 
Jefferson Lab~\cite{mjones,gayou}, together with data from SLAC
using Rosenbluth separation technique~\cite{slac}.
These new data~\cite{mjones,gayou} suggest 
that the proton Dirac ($F_1(Q^2)$) and Pauli form factor ($F_2(Q^2)$) scale
as ${Q\frac{F_1}{F_2}} \sim {\rm constant}$ at large values of $Q^2$.
Contributions from nonzero parton orbital angular momentum 
are power suppressed as shown by Lepage and Brodsky~\cite{lepage80}.
However, they are shown to lead to asymptotic scaling of the 
proton form factor ratio: 
$F_2(Q^2)/F_1(Q^2)\sim ({\rm log}^2Q^2/ \Lambda^2)/Q^2$ 
with 0.2 GeV$\le \Lambda \le $0.4 GeV based on an explicit 
pQCD calculation~\cite{belisky} or 
$F_2(Q^2)/F_1(Q^2)\sim 1/\sqrt{Q^2}$~\cite{ralston02,miller} 
that agrees with the JLab proton form factor data~\cite{mjones,gayou}.
A recent nonperturbative analysis~\cite{brodsky_ff} of the hadronic 
form factors based on light-front wave functions also describes the 
JLab proton form factor data~\cite{mjones,gayou} well.

While the intriguing $Q^2$ dependence of the proton form factor ratio can be
described~\cite{belisky,ralston02,miller,brodsky_ff}, it is important to 
understand the discrepancy between results obtained from recoil proton 
polarization measurements and those from Rosenbluth method.
New Jefferson Lab data~\cite{hallc} (solid circles in Fig.~1) 
from Rosenbluth separation are 
in good agreement with previous SLAC results. Recently, a new, 
``SuperRosenbluth'' experiment was carried out at 
Jefferson Lab~\cite{arrington}, in which the struck protons were detected 
to minimize systematic uncertainties associated with regular 
Rosenbluth technique in which scattered electron is detected. 
Preliminary results~\cite{arrington2} from 
the ``SuperRosenbluth'' experiment agree with previous Rosenbluth experiments.
Two-photon exchange contributions~\cite{wally} are
believed to contribute to the observed discrepancy between the polarization
method and the Rosenbluth technique. Currently, there are intensive efforts
both in theory~\cite{tpe_theory} and in experiment~\cite{tpe_exp} 
aiming at understanding the two-photon 
exchange contributions to electron scattering in general, particularly to the
aforementioned discrepancy in the proton form factor ratio. 
A new experiment~\cite{blast} in which longitudinally polarized
electrons scattering off a polarized proton target is currently ongoing at 
MIT-Bates and the proton electric to magnetic form factor ratio will be 
extracted with high precision up to a $Q^2$ value of about 0.8 (GeV/c)$^2$.
Such a double-polarization experiment is important because it employs 
a completely different experimental technique with different 
systematic uncertainties than recoil proton polarization measurements.

\section{Neutron Electromagnetic Form Factors}

Measurements of the neutron electric form factor are
extremely challenging because of the lack of free neutron targets, 
the smallness of the $G^n_E$, and the dominance of the magnetic contribution
to the unpolarized differential cross-section. A promising approach to measure
$G^n_E$ is by using polarization degrees of freedom. 
For coincidence elastic scattering of longitudinally polarized electrons 
from ``free'' neutrons, $n(\vec{e},e'\vec{n})$ process, 
the recoil neutron polarization ratio ${\frac{P_x}{P_z}}$ is sensitive
to the neutron electric to magnetic form factor ratio~\cite{arnold2}.
Experiments with longitudinally polarized electron beams and recoil 
neutron polarimeters have been carried out at MIT-Bates~\cite{eden} 
and Mainz~\cite{ostrick,herberg} in the relatively low $Q^2$ region,
and $G^{n}_{E}$ has been extracted from the $d(\vec{e},e'\vec{n})$ process,
using the state-of-the-art two-body calculations 
by Arenh\"{o}vel~\cite{arenhovel}.
Most recently, such an approach has been employed at Jefferson Lab up 
to a $Q^2$ value of 1.5 (GeV/c)$^2$~\cite{madey}.

Alternatively, one can employ a vector polarized deuteron target 
or a polarized $^3$He target to
probe the neutron electric form factor by the $\vec{d}(\vec{e},en)$
reaction or the  $\vec{^{3}He}(\vec{e},en)$ process. 
A polarized $^3$He nucleus is an effective neutron target because its
ground state is dominated by a spatially symmetric $S$
wave in which the proton spins cancel and the spin of the $^3$He
nucleus is carried by the unpaired neutron~\cite{BW84,frier90}. 
The spin-dependent asymmetries
from the $\vec{d}(\vec{e},en)$ reaction for vector polarized deuteron
and  from $\vec{^{3}He}(\vec{e},en)$ process
give access to the quantity ${\frac{G^n_E}{G^n_M}}$ to first order 
when the target spin direction is aligned perpendicular to 
the momentum transfer vector $\vec{q}$.
The neutron electric form factor was extracted for the first 
time~\cite{nikhef} from a $\vec{d}(\vec{e},e'n)$ measurement at NIKHEF
in which a 
vector polarized deuteron target from an atomic beam source was employed.
More recently, a $\vec{d}(\vec{e},e'n)$ experiment~\cite{zhu,warren}
using a dynamically polarized solid deuterated ammonia target was carried out
at Jefferson Lab and $G^n_E$ was extracted at $Q^2$ values of 0.5 and 1.0 
(GeV/c)$^2$.
Following the first measurement on $G^{n}_{E}$ from 
$^{3}\vec{\rm He}(\vec{e},e'n)$ at Mainz \cite{Meyerhoff}, two more 
experiments~\cite{becker,rohe} were carried out. All three experiments 
employed a high pressure polarized $^{3}$He target achieved by
the metastability-exchange optical pumping technique and the compression
method. 

\begin{figure}[htbp]
\includegraphics[width=3.5in]{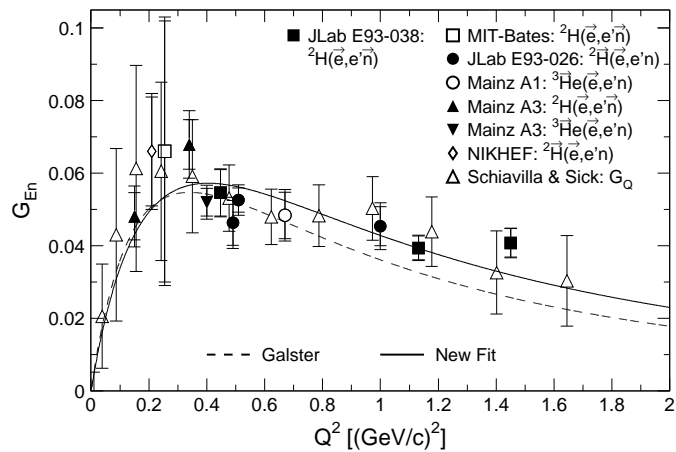}
\caption{\footnotesize\it Recent data on $G^n_E$ from polarization 
experiments. Also shown are the extracted $G^n_E$ values 
from the deuteron quadrupole form factor
data by Schivilla and Sick~\protect\cite{rocco}. The Galster 
parameterization~\cite{galster} as well as a new fit~\cite{madey}
are also shown.}
\label{fig:genpol}
\end{figure}

To extract $G^n_E$ information from these polarized target experiments, 
corrections for meson-exchange currents, final state interactions, etc. 
are necessary using the state-of-the-art two-body and three-body calculations.
Discussions on these corrections can be found in Ref. [30].
Fig.~\ref{fig:genpol} shows $G^n_E$ data as a function of $Q^2$ 
from polarization experiments. 
Also shown in Fig.~\ref{fig:genpol} are 
the extracted $G^n_E$ values from the deuteron quadrupole form factor
data by Schivilla and Sick~\cite{rocco}, and the Galster 
parameterization~\cite{galster}. New precision data~\cite{blast2} 
on $G^n_E$ in the low
$Q^2$ region will become available in the near future from MIT-Bates, and two 
approved experiments~\cite{jlab_gen} at Jefferson Lab will extend the measurement of $G^n_E$ to
much higher values of $Q^2$.

Until recently, most data on $G_M^n$ had been deduced from elastic and
quasi-elastic electron-deuteron scattering.  For inclusive
measurements, this procedure requires the separation of the longitudinal 
and transverse cross sections and the subsequent subtraction of a large
proton contribution. Thus, it suffers from large theoretical uncertainties
due in part to the deuteron model employed and in part to 
corrections for final-state
interactions (FSI) and meson-exchange currents (MEC).  
These complications can largely be avoided
if one measures the cross-section ratio of 
$d(e,e'n)$ to $d(e,e'p)$ at quasi-elastic kinematics.  Several recent
experiments \cite{Ankl94,Brui95,Ankl98,kubon2002} have employed this
technique to extract $G_M^n$ with uncertainties of 
$<$2\%~\cite{Ankl98,kubon2002} at $Q^2$ below 1 (GeV/c)$^2$.
Despite the high precision reported, however,
there is considerable disagreement among some of the
experiments~\cite{Mark93,Ankl94,Brui95,Ankl98,kubon2002} 
with respect to the absolute value of $G_M^n$.
The most recent deuterium data \cite{kubon2002} 
further emphasize this discrepancy. 
While the discrepancies among the deuterium experiments 
described above may be understood \cite{jordan},
additional data on $G_M^n$, preferably obtained using a complementary
method, are highly desirable.
Inclusive quasi-elastic $^3\vec{\rm He}(\vec{e},e')$ scattering
provides just such an alternative approach \cite{Gao94}.
Recently precision data on $G^n_M$ have been obtained from
inclusive quasi-elastic $^3\vec{\rm He}(\vec{e},e')$ process 
at Jefferson Lab 
\cite{xu,e95001_prc}. 
These new data are in very good agreement with the recent deuterium ratio 
measurements from Mainz \cite{Ankl98,kubon2002}, and in disagreement 
with results by Bruins {\it et al.} \cite{Brui95}. 
The deuterium ratio method was employed recently at Jefferson 
Lab~\cite{brook2} up to a $Q^2$ value of 4.7 (GeV/c)$^2$.

\begin{figure}[htbp]
\includegraphics[width=2.5in,angle=90]{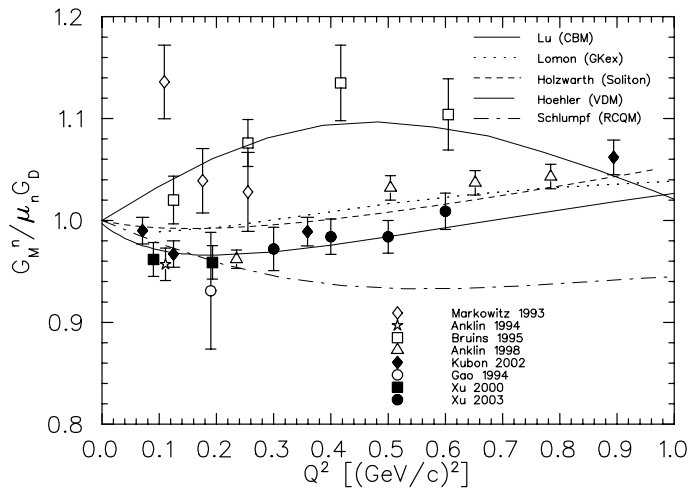}
\caption{{\footnotesize\it The neutron magnetic form factor $G_{M}^{n}$ data  
published since 1990, in units of the standard dipole form 
factor parameterization $G_D$, as a function of
$Q^{2}$. 
The Q$^2$ points of Anklin 94~\cite{Ankl94} and Gao 94~\cite{Gao94} 
have been shifted slightly for clarity. Also plotted are a few selected 
models of nucleon form factor calculation and the references are contained in
[45].}}
\label{fig:gmn}
\end{figure}

\section*{Acknowledgments}

I thank Eric Christy, Thia Keppel, Mark Jones, Dick Madey, Bradley Plaster, 
Andrei Semenov, and Wang Xu for providing helpful 
information about their experiments and for the preparation of
some of the figures. 
This work is supported   
by the U.S. Department of Energy under 
contract number DE-FC02-94ER40818 and DE-FG02-03ER41231. The author 
also acknowledges the OJI award in Nuclear Physics 
from the U.S. Department of Energy.

\appendix

\end{document}